\documentstyle[psfig,epsf,wrapft]{ptptex}
\newcommand{\bold}[1]{\mbox{\boldmath $#1$}}    % bold symbol
\newcommand{\comment}[1]{}			% Commented-Out Text
\title { {\it{SU}}(2) Chiral Sigma Model and Properties of Neutron Stars }
\author{Pradip Kumar {\sc Sahu}\thanks{JSPS fellow.}
	and
	Akira {\sc Ohnishi}
}
\inst{
Division of Physics, Graduate School of Science\\
Hokkaido University, Sapporo 060-0810, Japan \\
}
\recdate{
        July 12, 2000
}
\abst{
We discuss the {\it SU}(2) chiral sigma model in the context of nuclear 
matter using a mean field approach at high density.
In this model we include a dynamically generated isoscalar vector field
and higher-order terms in the scalar field.
With the inclusion of these, we reproduce the empirical values of the 
nuclear matter saturation density, binding energy, and nuclear 
incompressibility.
The value of the incompressibility is chosen according to 
recently obtained heavy-ion collision data.
We then apply the same dynamical model to neutron-rich matter in beta 
equilibrium, related to neutron star structure. 
The maximum mass and corresponding radius of stable non-rotating
neutron stars are found to be in the observational limit.
}

\begin{document}
\maketitle

\section{Introduction}
It has been argued that chiral symmetry is a good 
hadron symmetry~\cite{glen86} which ranks only below isotopic 
spin symmetry.
The spontaneous breaking of this symmetry generates the constituent quark
masses and hence various hadron masses, including the nucleon mass.
Therefore, the theories of dense nuclear matter, where dynamical 
modifications of hadron masses are expected should possess this 
symmetry.
In recent decades, three-body forces in the equation of state at 
high density have been studied by several authors.\cite{jack85,ains87}
%
%Also three-body forces support the importance of studying the chiral 
These studies also support the importance of studying the chiral 
sigma model, because the non-linear terms in the chiral sigma Lagrangian 
can give rise to three-body forces.
Theories of dense nuclear matter should be capable of 
describing the bulk properties of nuclear matter,
such as the binding energy per nucleon, saturation density, 
compression modulus/incompressibility and symmetry energy.
Presently there is no theory of dense nuclear matter 
that describes all the nuclear matter properties and possesses
chiral symmetry.
\par
A chiral Lagrangian using a scalar (sigma) field was originally introduced
by Gell-Mann and Levy,\cite{gell60} and later the importance of chiral 
symmetry in nuclear matter was emphasized by Lee and Wick.\cite{lee74}
The usual theory of pions does not possess the empirically desirable 
saturation properties for nuclear matter.
For this reason, an isoscalar vector field with a dynamically generated
mass was introduced \comment{via the Higgs mechanism}
into the theory of nuclear matter and it enabled us to
have a saturation density in nuclear matter equation of 
state.\cite{bogu83}
In the standard sigma model, the value of the incompressibility parameter
of nuclear matter turns out to be quite large
--- several times the desirable value ---
and can be reduced only 
by introducing the scalar field self-interactions with adjustable 
coefficients.
Several papers~\cite{prak87,glen88} have attempted to
derive the chiral sigma model equation of state at high matter 
density with normal nuclear matter saturation as well as a desirable
incompressibility value.
However, in these theories, the mass of the isoscalar vector field
is not generated dynamically. 
This fact can be considered a shortcoming of the chiral 
symmetry model.
\par
A few years ago, we proposed the {\it SU}(2) chiral sigma 
model~\cite{sahu93} 
to describe the properties of nuclear matter. In that work
we adopted an approach in which the mass of the isoscalar vector 
field is generated dynamically.
To ensure the saturation properties of nuclear matter, inclusion of 
such a vector field is necessary.
The nucleon effective mass thus acquires a self-consistent density 
dependence both on the scalar and the vector meson fields, and these meson 
fields are treated in the mean-field theory. 
To describe the nuclear matter properties, we have two parameters 
in the theory: the ratio of the coupling constants to the scalar
and to the isoscalar vector fields.
In this approach the value of incompressibility at saturation 
density is relatively large, which is an undesirable feature,
as far as nuclear matter at saturation and higher densities is 
concerned.
In the present calculation, we rectify the above mentioned 
shortcoming at the nuclear matter saturation density by including 
higher-order terms of the scalar field potential in our proposed chiral 
sigma Lagrangian.\cite{sahu93}
In this way, we get two extra parameters in the mean-field
approach. These are fitted with the nuclear matter properties
at the saturation density.
\par
The paper is organized as follows: In \S 2 we briefly describe 
the proposed {\it SU}(2) chiral sigma model and the derivation of 
the equation of state for the nuclear matter density.
Section 3 contains the neutron star matter equation of 
state in beta equilibrium with the neutron star results. 
A conclusion and summary are presented in \S 4. 

\section{ The {\it SU}(2) chiral sigma model and equation of state}
\par
The Lagrangian for the {\it SU}(2) chiral sigma model can be written as 
(we choose units in which $\hbar = c = 1$)
\par
\begin{eqnarray}
{\cal L} &=&  \frac{1}{2}\big(
	 \partial_{\mu} \bold{\pi} \cdot \partial^{\mu} \bold{\pi}
	+\partial_{\mu} \sigma \partial^{\mu} \sigma
	\big) 
- \frac{1}{4} F_{\mu\nu} F_{\mu\nu} 
\nonumber \\
&&- \frac{\lambda}{4}\big(x^2 - x^2_o\big)^2
- \frac{\lambda B}{6m^2}\big(x^2 - x^2_o\big)^3
- \frac{\lambda C}{8m^4}\big(x^2 - x^2_o\big)^4 
\nonumber \\
&&- g_{\sigma} \bar{\psi} \big(
		\sigma + i\gamma_5 \bold{\tau}\cdot\bold{\pi}
	\big) \psi 
+ \bar\psi \big(
i\gamma_{\mu}\partial^{\mu} - g_{\omega}\gamma_{\mu}
\omega^{\mu}\big) \psi 
\nonumber \\
&&+ \frac{1}{2}{g_{\omega}}^{2}
x^2 \omega_{\mu}
\omega^{\mu}
+ \frac{1}{24}\xi {g_w}^4(\omega_{\mu}\omega^{\mu})^2 -D\sigma,
\end{eqnarray}
where $F_{\mu\nu}\equiv\partial_{\mu}\omega_{\nu}-\partial_{\nu}
\omega_{\mu}$ and  $x^2= \bold{\pi}^2+\sigma^{2}$, $\psi$ is the nucleon 
isospin doublet, $\bold{\pi}$ is the pseudoscalar-isovector pion field,
$\sigma$ is the scalar field, and $D$ is a constant.
The Lagrangian includes a dynamically generated isoscalar vector field, 
$\omega_{\mu}$, that couples to the conserved baryonic current 
$j_{\mu}=\bar{\psi}\gamma_{\mu}\psi$.
$B$ and $C$ are constant parameters included the higher-order 
self-interaction of the scalar field in the potential.
In the fourth order term in the omega fields, the quantity $\xi$ 
is a constant parameter.
For simplicity, we set $\xi$ to zero in our calculation.
\par
The interactions of the scalar and the pseudoscalar mesons with the
vector meson generate a mass of the latter 
through the spontaneous breaking of the chiral symmetry.
The masses of the nucleon, the scalar meson and the vector meson
are respectively given by 
\begin{eqnarray}
M = g_{\sigma} x_o,~~ m_{\sigma} = \sqrt{2\lambda} x_o,~~
m_{\omega} = g_{\omega} x_o\ ,
\end{eqnarray}
where $x_o$ is the vacuum expectation value of the $\sigma$ field,
$\lambda~=~({m_{\sigma}}^{2}-{m_{\pi}}^{2})/(2 {f_{\pi}}^{2})$, with
$m_{\pi}$ the pion mass and $f_{\pi}$ the pion decay coupling
constant, and $g_{\omega}$ and $g_{\sigma}$ are the coupling constants 
for the vector and scalar fields, respectively.
Throughout this paper, we limit ourselves to a mean field treatment
and ignore the explicit role of $\pi$ mesons.
\par
The equation of motion of fields is obtained by adopting the mean-field 
approximation. 
This approach has been used extensively to obtain field theoretical 
equation of state models for high density matter.
Using the mean-field ansatz, the equation of motion for the isoscalar 
vector field is

\begin{equation}
\omega_0=\frac{n_B}{g_{\omega} x^2}\ ,
\end{equation}
and the equation of motion for the scalar field in terms
of $y \equiv x/x_o$ is

\begin{eqnarray}
(1-y^2) -\frac{B}{m^2 c_{\omega}}(1-y^2)^2
+\frac{C}{m^4c_{\omega}^2}(1-y^2)^3 \nonumber \\
+\frac{2 c_{\sigma}c_{\omega}n_B^2}{M^2y^4}
-\frac{c_{\sigma}\gamma}{\pi^2}\int^{k_F}_o\frac{k^2dk}
	{\sqrt{k^2+M^{\star 2}}}=0\ ,
\end{eqnarray}
where $M^{\star} \equiv yM$ is the effective mass of the nucleon and
\begin{equation}
c_\sigma \equiv  g_{\sigma}^2/m_{\sigma}^2\ , \quad\quad\quad
  c_{\omega} \equiv g_{\omega}^2/m_{\omega}^2\ .
\end{equation}
The quantity $n_B$ is the baryon density, or equivalently,
$n_B=\frac{\gamma}{(2\pi)^3}\int^{k_F}_o d^3k$,
where $k_F$ is the Fermi momentum and $\gamma$ is the nucleon spin 
degeneracy factor.
\par
The equation of state is calculated from the diagonal components of the 
conserved total stress tensor corresponding to the Lagrangian together 
with the mean-field equation of motion for the fermion field and a
mean-field approximation for the meson fields.
The total energy density, $\varepsilon$, and pressure, $P$, of 
the many-nucleon system are the following:
\begin{eqnarray}
\varepsilon
&=&
	  \frac{M^2(1-y^2)^2}{8c_{\sigma}}
	- \frac{B}{12c_{\omega}c_{\sigma}}(1-y^2)^3
	+ \frac{C}{16m^2c_{\omega}^2c_{\sigma}}(1-y^2)^4
\nonumber \\
&&+
	  \frac{c_{\omega} n_B^2}{2y^2} 
	+ \frac{\gamma}{2\pi^2}
		\int _o^{k_F} k^2dk\sqrt{{k}^2 + M^{\star 2}}\ ,
\nonumber\\
P &=&
	- \frac{M^2(1-y^2)^2}{8c_{\sigma}}
	+ \frac{B}{12c_{\omega}c_{\sigma}}(1-y^2)^3
	- \frac{C}{16m^2c_{\omega}^2c_{\sigma}}(1-y^2)^4
\nonumber \\
&&+
	  \frac{c_{\omega}n_B^2}{2y^2} 
	+ \frac{\gamma }{6\pi^2}
		\int _o^{k_F} \frac{k^4dk}{\sqrt{{k}^2 + M^{\star 2}}}\ .
\end{eqnarray}
\begin{wrapfigure}{r}{6.6cm}
\psfig{figure=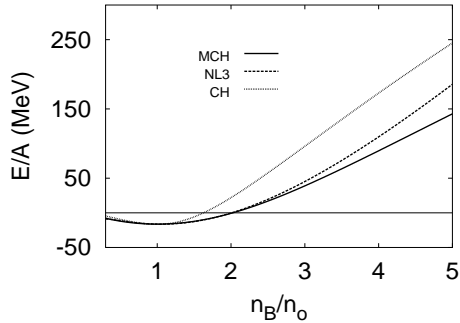,width=6.6cm}
\caption{
Energy per nucleon as a function of baryon density in units of $n_0$.
The solid curve (MCH) corresponds to modified chiral sigma model,
where the nuclear incompressibility is around 300 MeV, the dashed
curve (NL3) corresponds to the equation of state derived from recent 
heavy-ion collision data~\cite{sahuj00} with incompressibility around 
340 MeV, and 
the dotted curve corresponds to the original chiral sigma 
model,\cite{sahu93} with very high incompressibility.
}
\end{wrapfigure}
The energy per nucleon is $E/A=\varepsilon/n_B$,
where $\gamma=4$ for symmetric nuclear matter.
\par
In the above equations, we have four parameters:
the nucleon coupling to the scalar and the vector fields,
$c_{\sigma}$ and $c_{\omega}$,
%the nucleon coupling to the vector field $c_{\omega}$ 
and the coefficients in the scalar potential terms, $B$ and $C$. 
These are obtained by fitting the saturation values of binding 
energy/nucleon ($-16.3$ MeV), baryon density ($0.153~\hbox{fm}^{-3}$),
and effective (Landau) mass ($0.85M$).\cite{moll88}
The nuclear incompressibility is somewhat uncertain at saturation
and therefore we choose it between $250$ MeV and $350$ MeV,
i.e. $\sim 300$ MeV, in accordance with
recent heavy-ion collision data.\cite{sahu98,sahu00}
These values are $c_{\omega}=1.9989~ \hbox{fm}^2$,
$c_{\sigma}=6.8158~ \hbox{fm}^2$,
$B=-100$ and  $C=-133.6$.
In Fig.~1, we present the energy/nucleon as a function of  the baryon 
density.
The solid curve (MCH) corresponds to the modified chiral sigma model
presented above, where the nuclear incompressibility is around 300 MeV.
This equation of state is much softer than that of the original chiral
sigma model (CH), which is represented by the dotted curve.
This is due to the additional terms $B$ and $C$ in the modified chiral 
sigma model in the potential term in equation (1).
For comparison, we display the dashed curve (NL3), which
was derived from recent heavy-ion collision data~\cite{sahuj00}
with incompressibility around 340 MeV.
We notice from Fig.~1 that the MCH model is softer than the NL3 model at 
high density. This is due to the slight difference in the incompressibility.
\section{Equation of state in neutron star matter}
\par
In the interior of neutron stars, i.e. at high density, the neutron
chemical potential exceeds the combined mass of the proton and electron.
Therefore, asymmetric matter, with an admixture of electrons,
rather than pure neutron matter, is a more likely composition of matter
in neutron star interiors.
The concentrations of neutrons, protons and electrons can be determined
from the condition of beta equilibrium ($n\leftrightarrow p+e+\bar{\nu}$)
and from charge neutrality, assuming that neutrinos are not degenerate.
We have 
\begin{eqnarray}
\mu_n = \mu_p + \mu_e,~~ n_p = n_e, 
\end{eqnarray}
where $\mu_i$ is the chemical potential of particle species $i$.
For the purpose of describing neutron-rich matter, we include the 
interaction due to the isospin triplet $\rho$ meson in the Lagrangian (1).
The following terms are included in the Lagrangian:

\begin{equation}
-\frac {1}{4}\bold{G}_{\mu\nu}\cdot\bold{G}^{\mu\nu}
+\frac{1}{2}m^2_{\rho}\bold{\rho}_{\mu}\cdot\bold{\rho}^{\mu}
-\frac{1}{2}g_{\rho}\bar\psi
		(\bold{\rho}_{\mu}\cdot\bold{\tau}\gamma^{\mu})
		\psi\ .
\end{equation}
Here
$\bold{G}_{\mu\nu} \equiv \partial_{\mu}\bold{\rho}_{\nu}-\partial_{\nu}
\bold{\rho}_{\mu}$.
Using the mean-field approximation in the equation of motion for
$\rho$, the following density dependence equation is obtained:

\begin{equation}
\rho^3_o = \frac{g_{\rho}}{2m_\rho^2} (n_p-n_n)\ .
\end{equation}
From the semi-empirical nuclear mass formula, the symmetric energy
coefficient is
\begin{equation}
a_{\rm sym} = \frac{c_{\rho} k_F^3}{12\pi^2} + \frac{k_F^2}{6\sqrt{(k_F^2
+M^{\star 2})}}\ ,
\end{equation}
where $c_{\rho} \equiv g^2_\rho/m^2_{\rho}$ and $k_F=(6\pi^2n_B/\gamma)
^{1/3} (n_B=n_p+n_n)$.
We fix the coupling constant $c_{\rho}$ by requiring that $a_{\rm sym}$ 
correspond to the empirical value, 32 MeV.\cite{moll88}
This gives $c_{\rho}=4.66~ \hbox{fm}^2$.
The inclusion of the $\rho$ meson in the Lagrangian will contribute the term
$m^2_{\rho}(\rho_o^3)^2/2$ to the energy density and pressure.
Then the equation of state for neutron rich nuclear matter in beta 
equilibrium is calculated using the expression for
the energy density $\varepsilon$ and the pressure $P$
as follows:

\begin{eqnarray}
\varepsilon &=& \frac{M^2(1-y^2)^2}{8c_{\sigma}}
-\frac{B}{12c_{\omega}c_{\sigma}}(1-y^2)^3
+\frac{c_{\omega} n_B^2}{2 y^2} 
\nonumber \\
&&+\frac{C}{16m^2c_{\omega}^2c_{\sigma}}(1-y^2)^4
+ \sum_i \varepsilon_{FG} + \frac{1}{2}m_{\rho}^2 {\rho_0}^2\ ,
\nonumber\\
P &=& - \frac{M^2(1-y^2)^2}{8c_{\sigma}}
+\frac{B}{12c_{\omega}c_{\sigma}}(1-y^2)^3
+ \frac{c_{\omega} n_B^2}{2 y^2} 
\nonumber \\
&&-\frac{C}{16m^2c_{\omega}^2c_{\sigma}}(1-y^2)^4
+\sum_i P_{FG} + \frac{1}{2}m_{\rho}^2 {\rho_0}^2\ .
\end{eqnarray}
In these equations $\varepsilon_{FG}$ and $P_{FG}$ are the relativistic
non-interacting energy density and pressure of the baryons (with effective
masses) and electrons ($i$), respectively.
\par
Figure 2, displays the pressure versus the total mass-energy density
for neutron-rich matter in beta equilibrium.
The solid curve corresponds to the modified chiral sigma model MCH 
considered in the present calculation.
Inclusion of the two parameters, $B$ and $C$ in the potential term,
gives a reasonable value for the incompressibility at the saturation 
density of nuclear matter.
Hence, the MCH model is much softer than the original CH model
with regard to the neutron star matter equation of state.
The physics behind this is that the pressure generated by the
self-interaction of scalar fields at high density is less;
i.e., the pressure with density decreases more than that for
the CH model.
The dashed curve (NL3) represents the neutron rich equation of
state, which was derived from the heavy-ion collision 
data.\cite{sahuj00}
From Fig.~ 2, we note, in comparison with the NL3 model, that the 
MCH model is
stiffer in the low density range and becomes softer above three times 
the normal nuclear matter density.
\par
The mass and radius of a neutron star are characterized
by its structure. 
These are determined from the equations that describe the hydrostatic
equilibrium of degenerate stars without rotation in general relativity, 
the Tolman-Oppenheimer-Volkoff (TOV) equations:~\cite{misn70,hart67}
\begin{eqnarray}
\frac {dp}{dr} = - \frac{G (\epsilon + {{p/c^2}}) 
(m + {{4\pi r^3 p/c^2}})} {r^2 (1-{{2 Gm/rc^2}})}\ ,~~~
\frac {dm}{dr} = 4\pi r^2\epsilon\ .
\end{eqnarray}
\begin{wrapfigure}{r}{6.6cm}
\psfig{figure=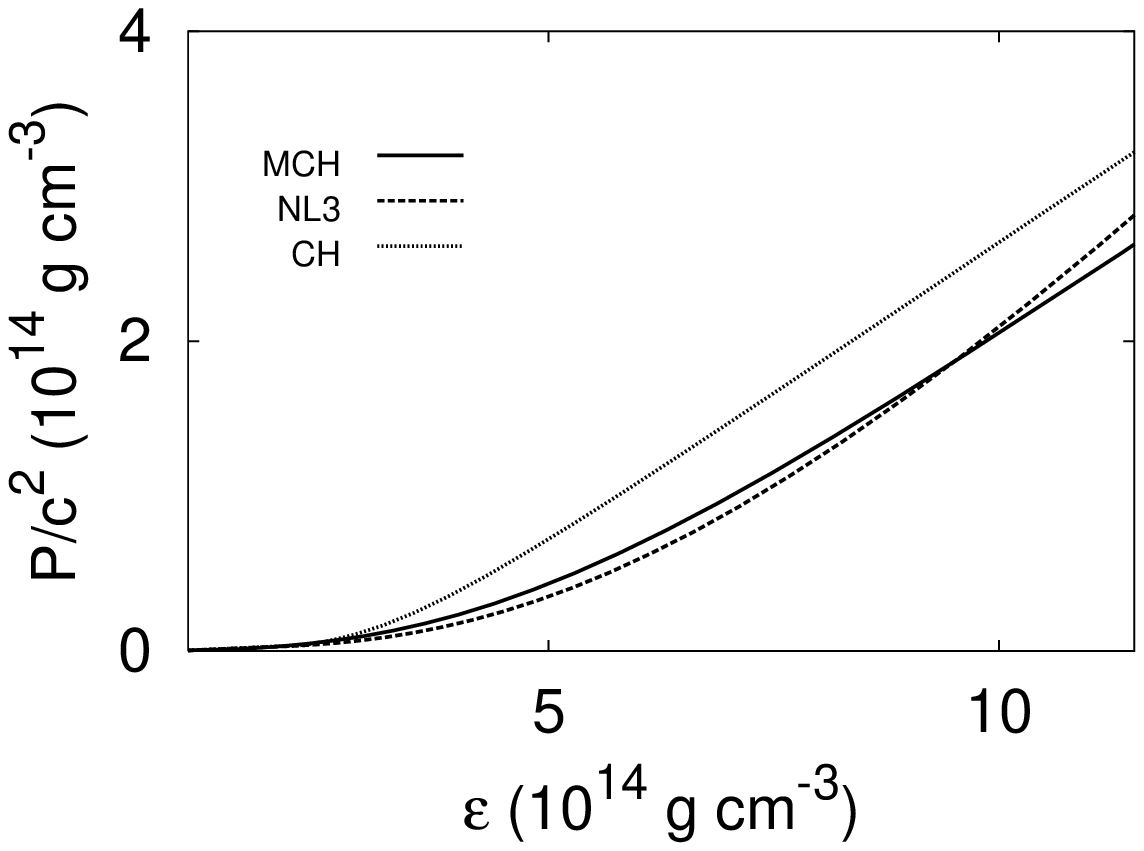,width=6.6cm}
\caption{The neutron star matter pressure as a function of the energy 
density. The models are the same as in Fig.~1. }
%\end{wrapfigure}
%
%\begin{wrapfigure}{r}{6.6cm}
%\psfig{figure=fig3.ps,height=6.5cm,width=6.6cm}
\psfig{figure=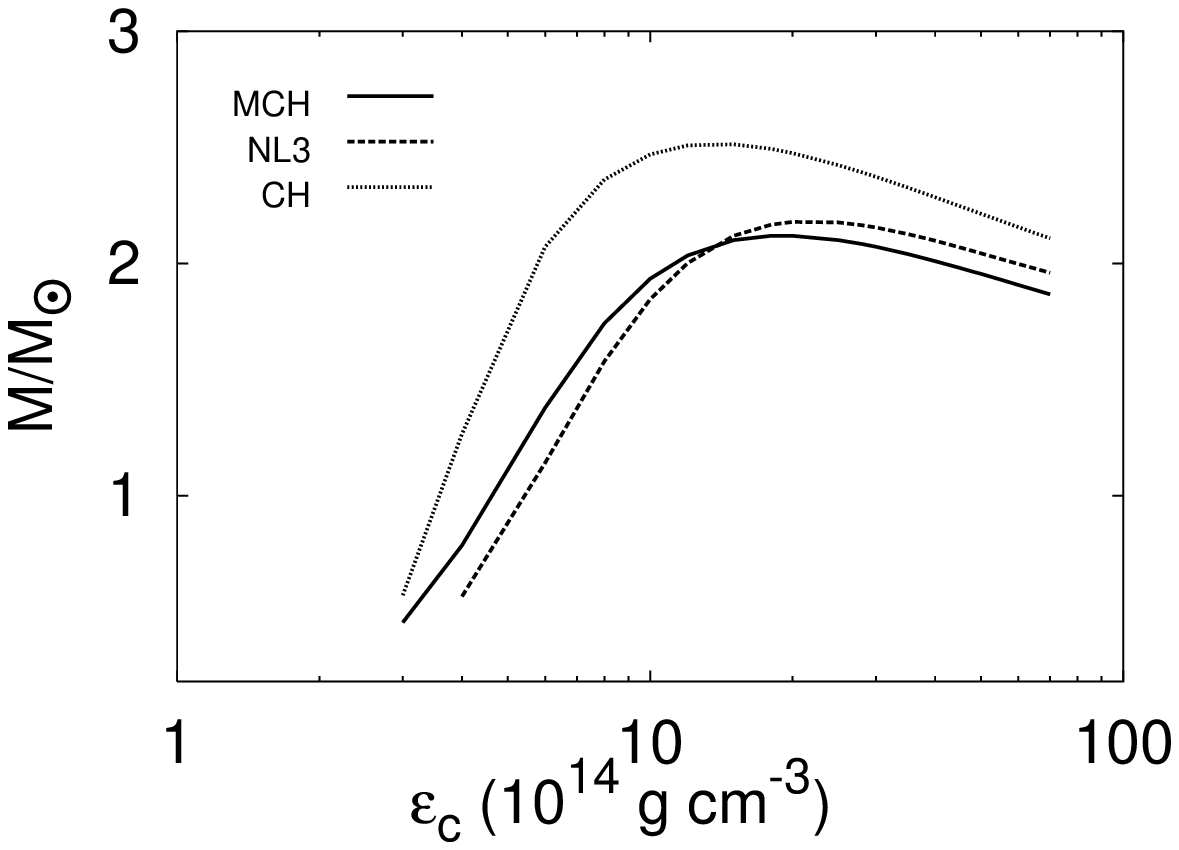,width=6.6cm}
\caption{The neutron star mass as a function radius. The models are
the same as in Fig.~1.}
\end{wrapfigure}
Here $p$ and $\epsilon$ are the pressure and total mass-energy density,
and $m(r)$ is the mass contained in a volume of radius $r$.
The quantity $G$ is the gravitational constant, and $c$ is the velocity of
light. 
To integrate the TOV equations, one needs to know the equation of state
for the entire expected density range of the neutron star, starting from
the high density at the center to the surface densities.
Therefore, we construct a composite equation of state for the entire
neutron star density span by joining our equation of state for
high density neutron rich matter to those with (i) $10^{14}$ to 
$5 \times 10^{10}~$g$~ \hbox{cm}^{-3}$,\cite{negl73} (ii) 
$5 \times 10^{10}~$g$ ~\hbox{cm}^{-3}$ to 
$10^3~$g$ ~\hbox{cm}^{-3}$,\cite{baym71}
and 
(iii) less than $10^3~$g$ ~\hbox{cm}^{-3}$.\cite{feyn49}
Thus we integrate the TOV equations for the newly constructed equation 
of state 
and given central density $\epsilon(r=0)=\epsilon_c$ with the boundary
condition $m(r=0)=0$ to give $R$ and $M$. 
The radius $R$ is defined by the point where $P\sim 0$, or, equivalently,
$\epsilon=\epsilon_s$, where $\epsilon_s$ ($7.8 $g$~\hbox{cm}^{-3}$) is the 
density expected at the star surface. 
The total mass is then given by $M=m(R)$.
\par
The results for the star structure parameters are listed in Table I and
displayed in Fig. 3.
This figure plots the mass as a function of the central density.
The models are the same as in Fig.~2.
\vskip 0.1 in
\begin{table}
\caption{Neutron star structure parameters.}
\vskip 0.1 in
\begin{center}
\begin{tabular}{ccccccccc}
\hline
\hline
\multicolumn{1}{c}{$\varepsilon_c$}&
\multicolumn{1}{c}{$R$} &
\multicolumn{1}{c}{$M/M_{\odot}$} &
\multicolumn{1}{c}{$z$} &
\multicolumn{1}{c}{$I$} &
\multicolumn{1}{c}{} \\
\multicolumn{1}{c}{($g~\hbox{cm}^{-3}$) } &
\multicolumn{1}{c}{($\hbox{km}$)} &
\multicolumn{1}{c}{} &
\multicolumn{1}{c}{}&
\multicolumn{1}{c}{(g $~\hbox{cm}^2$)} &
\multicolumn{1}{c}{}\\
\hline
2.0$\times 10^{14}$ &18.33&0.19&0.015&0.11$\times 10^{45}$ & \\
6.0$\times 10^{14}$ &13.77&1.38&0.19&1.77$\times 10^{45}$ & \\
8.0$\times 10^{14}$ &13.57&1.74&0.27&2.38$\times 10^{45}$ & \\
1.0$\times 10^{14}$ &13.28&1.93&0.32&2.66$\times 10^{45}$ & \\
1.5$\times 10^{15}$ &12.62&2.10&0.40&2.73$\times 10^{45}$ & MCH \\
1.8$\times 10^{15}$ &12.30&2.12&0.43&2.64$\times 10^{45}$ & \\
2.0$\times 10^{15}$ &12.11&2.12&0.44&2.57$\times 10^{45}$ & \\
2.5$\times 10^{15}$ &11.74&2.10&0.46&2.39$\times 10^{45}$ & \\
3.0$\times 10^{15}$ &11.44&2.07&0.47&2.24$\times 10^{45}$ & \\
\hline
2.0$\times 10^{14}$ &18.27&0.20&0.016&0.13$\times 10^{45}$ & \\
6.0$\times 10^{14}$ &13.08&1.14&0.16&1.24$\times 10^{45}$ & \\
8.0$\times 10^{14}$ &12.98&1.58&0.25&1.92$\times 10^{45}$ & \\
1.0$\times 10^{14}$ &12.75&1.85&0.32&2.34$\times 10^{45}$ & \\
1.5$\times 10^{15}$ &12.09&2.12&0.44&2.63$\times 10^{45}$ & NL3 \\
1.8$\times 10^{15}$ &11.76&2.17&0.48&2.61$\times 10^{45}$ & \\
2.0$\times 10^{15}$ &11.56&2.18&0.50&2.56$\times 10^{45}$ & \\
2.5$\times 10^{15}$ &11.15&2.17&0.54&2.41$\times 10^{45}$ & \\
3.0$\times 10^{15}$ &10.83&2.15&0.56&2.26$\times 10^{45}$ & \\
\hline
2.0$\times 10^{14}$ &23.68&0.13&0.008&0.07$\times 10^{45}$ & \\
6.0$\times 10^{14}$ &14.97&2.07&0.30&3.73$\times 10^{45}$ & \\
8.0$\times 10^{14}$ &14.76&2.36&0.38&4.36$\times 10^{45}$ & \\
1.0$\times 10^{14}$ &14.45&2.47&0.42&4.48$\times 10^{45}$ & \\
1.5$\times 10^{15}$ &13.77&2.51&0.47&4.19$\times 10^{45}$ & CH \\
1.8$\times 10^{15}$ &13.45&2.49&0.49&3.96$\times 10^{45}$ & \\
2.0$\times 10^{15}$ &13.27&2.48&0.49&3.81$\times 10^{45}$ & \\
2.5$\times 10^{15}$ &12.89&2.42&0.50&3.49$\times 10^{45}$ & \\
3.0$\times 10^{15}$ &12.60&2.37&0.50&3.23$\times 10^{45}$ & \\
\hline
\end{tabular}
\end{center}
\end{table}
In Table I, we also list additional parameters of interest. These
are the moment  of inertia $I$, and the surface redshift $z=\frac{1}
{\sqrt{1-2GM/Rc^2}} -1$ as a function of the central density of the star.
(For details, see Ref.~\citen{sahu93})
These are important for the dynamics and transport properties of pulsars.
From Fig. 3 and Table I, we see that the maximum masses of the stable
neutron stars are 2.1$M_{\odot}$, 2.2$M_{\odot}$ and 2.5$M_{\odot}$ and
corresponding radii are 12.1 km, 11.6 km and 13.8 km for MCH, NL3 and CH
equation of states, respectively. 
The corresponding central densities are $2.0\times 10^{15}~ $g$~ 
{\rm cm}^{-3}$ ($>$ 7 times nuclear matter density, where the 
nuclear matter density is $2.8\times 10^{14}~ $g$~ {\rm cm}^{-3}$), 
$2.0\times 10^{15}~ $g$~ {\rm cm}^{-3}$
($>$ 7 times nuclear matter density) and $1.5\times 10^{15}~ $g$~ 
{\rm cm}^{-3}$ ($>$ 5 times nuclear matter density) for MCH, NL3 and CH,
respectively, at the maximum neutron star masses.
From the neutron star structure point of view, our present model MCH is
comparable to that of the NL3 model, which was constructed using
recent heavy-ion collision data.\cite{sahuj00}
The maximum masses calculated using our models are in the range of recent
observations,\cite{para98,barz99,oros99,mill98} where the observational
consequences are given below.
\par
Very recently, it has been observed that the best determined neutron star
masses~\cite{thor99} are found in binary pulsars, and they all lie in the 
range 1.35$\pm 0.04M_{\odot}$, except for those of the non-relativistic 
pulsars 
PSR J1012+5307, for which $M=(2.1\pm 0.8)M_{\odot}$.\cite{para98}
There are several measured X-ray binary masses, and the heaviest 
among them are Vela X-1 with $M= (1.9\pm 0.2)M_{\odot}$~\cite{barz99}
and Cygnus X-2 with $M= (1.8\pm 0.4)M_{\odot}$.\cite{oros99}
From the recent discovery of high-frequency brightness oscillations in
low-mass X-ray binaries, the large masses of the neutron stars
are around $M=2.3M_{\odot}$, $M=2.1M_{\odot}$ and $M=1.9M_{\odot}$
in QPO 4U 1820-30, QPO 4U 1608-52 and QPO 4U 1636-536, 
respectively.\cite{mill98}
This provides a new method to determine the masses and radii 
of the neutron stars.
Our results lie in the range of those predicted by the observational 
limits.\cite{para98,barz99,oros99,mill98}
\section{Summary and conclusion}
We have discussed the {\it SU}(2) chiral sigma model, in which the 
isoscalar vector field is generated dynamically.
In the present calculation, we have modified the {\it SU}(2) chiral
sigma model by including two extra terms in the Lagrangian
to ensure an appropriate value of the incompressibility at the
saturation density of symmetric nuclear matter.
By employing these and using a mean field approximation,
we have obtained the nuclear equation of state at high densities.
This is compatible with the equation of state that was derived 
using recent heavy-ion collision flow data.\cite{sahu98,sahu00,sahuj00}
The nuclear equation of state is reasonably softer than
the original {\it SU}(2) chiral sigma model we considered 
previously.\cite{sahu93}
We then employed the same nuclear equation of state in the 
neutron star matter calculation, where the composition of the star 
matter consists of neutrons, protons and electrons.
The mass and radius of the neutron star were calculated from
the TOV equations, and were found to be  2.1$M_{\odot}$ and 12.1 km, 
respectively.
The corresponding central density is $2.0\times 10^{15}~ $g$~ 
{\rm cm}^{-3}$ ($>$ 7 times the nuclear matter density, 
$2.8\times 10^{14}~ $g$~ {\rm cm}^{-3}$) for the modified 
chiral sigma model.
These values are comparable to the values calculated using the recent 
NL3 model, which was recently constructed using heavy-ion collision 
data.
The overall structural values are in good agreement with those given by 
the quasi-periodic
oscillation method,\cite{mill98} a new method to determine the masses and 
radii of neutron stars.
\par
In the future we plan to investigate more systematically the properties of
neutron stars in the presence of other baryons, i.e., hyperons
and mesons with (hidden-)strangeness
in the {\it SU}(3) chiral sigma model with the proper
interactions between them derived from recent experiments.
Moreover, we are interested in searching for a phase transition to 
quark matter (a new state of matter) by implementing this equation 
of state in heavy-ion collision simulation calculations. 

\section*{Acknowledgements}

PKS would like to acknowledge the support of the Japan Society for the
Promotion of Science (ID No. P 98357), Japan.

\end{document}